\documentclass[aps,prb,reprint,twocolumn,amsmath,amssymb,citeautoscript,longbibliography]{revtex4-2}
\usepackage{dcolumn}
\usepackage{bm}
\usepackage{latexsym,epsfig}
\usepackage{graphicx}
\usepackage{verbatim}
\usepackage{comment}
\usepackage{amsmath}
\usepackage{amssymb}
\usepackage{physics}
\usepackage{stmaryrd}
\usepackage{color}
\usepackage{epstopdf}
\usepackage{grffile}
\usepackage{lipsum}
\usepackage{enumitem}
\usepackage{mathtools}
\usepackage[normalem]{ulem}
\usepackage{circuitikz}
\usepackage{soul}
\usepackage[dvipsnames]{xcolor}
\DeclareGraphicsExtensions{.eps}

\newcommand{\beq}{\begin{equation}}
\newcommand{\eeq}{\end{equation}}
\newcommand{\bea}{\begin{eqnarray}}
\newcommand{\eea}{\end{eqnarray}}
\newcommand{\ben}{\begin{eqnarray*}}
\newcommand{\een}{\end{eqnarray*}}
\newcommand{\bfig}{\begin{figure}}
\newcommand{\efig}{\end{figure}}

\usepackage{hyperref}
\hypersetup{
    colorlinks=true,      
    urlcolor=blue,
    citecolor=blue,
    linkcolor=blue
}

\usepackage{booktabs}
\usepackage{array}
\usepackage{lipsum} 
\usepackage{array} 

\begin{document}
\title{Engineering edge states in two-leg SSH ladder and their topoelectric circuit realization}
\author{Anish Kuanr}
\email{anish.kuanr@niser.ac.in }
\affiliation{School of Physical Sciences, National Institute of Science Education and Research, Jatni 752050, India}
\affiliation{Homi Bhabha National Institute, Training School Complex, Anushaktinagar, Mumbai 400094, India}
\author{Rajashri Parida}
\email{rajashriparida33@gmail.com }
\affiliation{School of Physical Sciences, National Institute of Science Education and Research, Jatni 752050, India}
\affiliation{Homi Bhabha National Institute, Training School Complex, Anushaktinagar, Mumbai 400094, India}
\author{Prabhu Prasad Tripathy}
\email{prabhup.tripathy@niser.ac.in }
\affiliation{School of Physical Sciences, National Institute of Science Education and Research, Jatni 752050, India}
\affiliation{Homi Bhabha National Institute, Training School Complex, Anushaktinagar, Mumbai 400094, India}
\author{Saralasrita Mohanty}
\email{saralasrita@niser.ac.in }
\affiliation{School of Physical Sciences, National Institute of Science Education and Research, Jatni 752050, India}
\affiliation{Homi Bhabha National Institute, Training School Complex, Anushaktinagar, Mumbai 400094, India}
\author{Tapan Mishra}
\email{mishratapan@gmail.com}
\affiliation{School of Physical Sciences, National Institute of Science Education and Research, Jatni 752050, India}
\affiliation{Homi Bhabha National Institute, Training School Complex, Anushaktinagar, Mumbai 400094, India}
\date{\today}

\begin{abstract}
We study the topological phase transition in a two-leg Su-Schrieffer-Heeger (SSH) ladder by redefining the unit-cell structure. For both identical hopping dimerization pattern (uniform) and alternate hopping dimerization pattern (staggered) along the legs of the ladder, we demonstrate that different unit-cell choices generate different topological phases and phase transitions. In the uniformly dimerized ladder, variation of the inter-leg coupling induces a transition from topological phase to another topological phase through a gapless region. In contrast, the staggered dimerization configuration exhibits a richer phase structure, supporting both topological-topological and trivial-topological transitions occurring through a single gap-closing point, depending on the unit-cell definition.  The phases are characterized through bulk-boundary correspondence, edge-state analysis, and bulk topological invariants. Interestingly, we obtain that while all the topological phases host two zero energy edge states each, the topological phase for the staggerred dimerization case at small inter-leg coupling hosts four edge states. We then perform topoelectric circuit simulation and experiments to observe the signatures of the topological phases. By using circuit impedance and voltage responses we establish the emergence of distinct edge modes in the circuit. Our analysis provides a route to engineer topological edge modes in a two-leg  SSH ladder set-up. 
\end{abstract}

\maketitle
\section{Introduction}
The concept of topology has profoundly reshaped our understanding of quantum phases of matter by revealing states that cannot be characterized within the traditional Landau paradigm of symmetry breaking and local order parameters~\cite{TKNN,Haldane1988, Hassanreview}. 
Instead, topological phases are distinguished by global invariants leading to robust boundary modes that persist against local perturbations as long as certain symmetries are preserved~\cite{Chen2012, Pollmann2012, senthil_review}.
These features are collectively captured under the notion of symmetry-protected topological (SPT) phases, which have become a central theme in modern condensed matter physics due to their simplicity in terms of short-range entanglement structure, conceptual richness, and technological potential~\cite{Hassanreview, senthil_review}.
Among the simplest and most instructive models realizing an SPT phase is the one-dimensional Su-Schrieffer-Heeger (SSH) model~\cite{ssh_model}, where dimerized nearest-neighbor hopping amplitudes give rise to a topological phase transition from a trivial insulating phase as the hopping dimerization is appropriately varied. 
Despite its minimal structure, the SSH model has served as a versatile platform for exploring fundamental aspects of topology, including bulk-boundary correspondence, quantized geometric phases, topological transport, and topological phase transitions. 
Its conceptual simplicity has enabled direct experimental realizations across a wide range of systems, including ultracold atoms~\cite{Atala2013,Takahashi2016pumping,Lohse2016,ssh_expt_2,ssh4_expt,Leder2016}, Rydberg atoms~\cite{browyes,rydberg_atom_review}, photonic and acoustic lattices~\cite{Mukherjee2017,Lu2014,Kitagawa2012,seba_soliton}, mechanical metamaterials~\cite{ssh_expt_1}, superconducting platforms~\cite{Liu-superconducting}, semiconductor nanolattices~\cite{semicoductor_nanolattice}, and, more recently, electrical circuits~\cite{EC1, EC2}.

A richer landscape emerges when moving beyond strictly one-dimensional chains to quasi-one-dimensional geometries such as ladders. 
Ladder systems provide a natural setting to investigate the dimensional crossover by interpolating between one and two dimensions and host new physical mechanisms stemming from inter-chain coupling and proximity effects. 
The linkage of two distinct SSH chains introduces unique geometric and crystalline symmetries, including glide reflection symmetry~\cite{glide_reflection}. These  symmetries give rise to unconventional quantum phenomena, such as complex band structures and charge fractionalization.

The simplest set-up in the context is the two coupled SSH chains or SSH ladder which has been shown to exhibit novel topological scenarios in recent years.
Recent studies have shown that when two identical SSH chains of topological configuration are coupled with each other, the inter-chain coupling immediately destroys the topological nature of the system~\cite{padhan_ladder}. However, when one topological and one trivial chain are coupled, then there occurs a topological to trivial phase transition as a function of the inter-leg coupling. It has also been shown that when an SSH chain of either topological or trivial character is coupled with a chain of uniform hopping, a topological phase can be induced by the inter-chain hopping. These emergent topological phase transitions reveal the peculiar role of inter-leg couplings in ladder systems~\cite{parida_coupling}. It is to be noted that the onset of topological phases in such coupled SSH chains are found to be due to the emergence of effective SSH-type chains which satisfies either topological or trivial configuration, depending on the choice of hopping strengths involved. 
\begin{figure*}[t]
\begin{center}
\includegraphics[width=2\columnwidth]{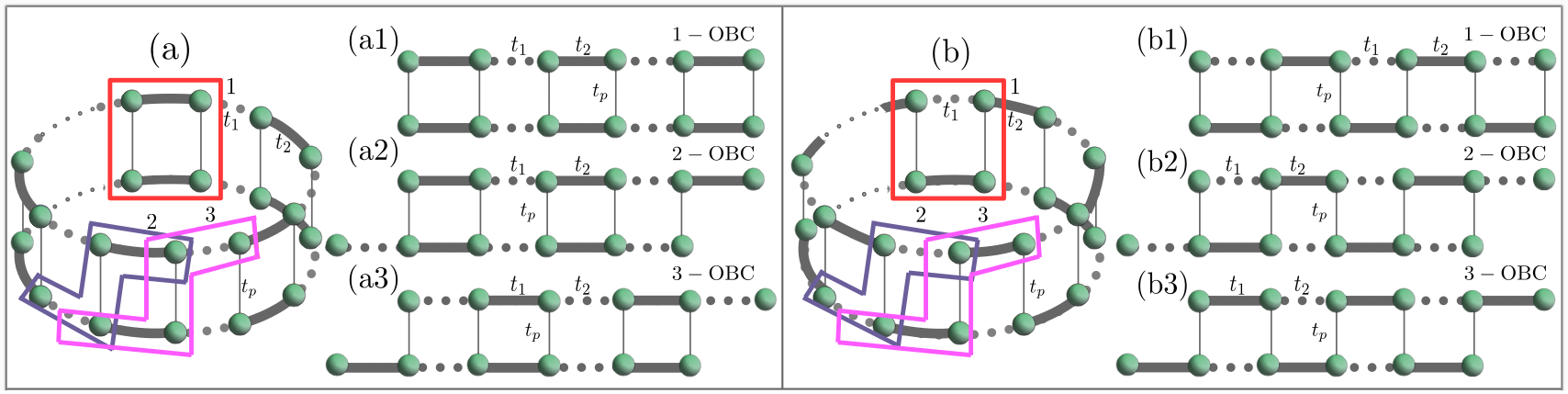}
\end{center}
\caption{Schematic of two types of ladder configurations. The left (right) panel indicates the ladder in uniform (staggered) configuration. The red, blue, and magenta boxes highlight three possible unit cell configurations in both cases, which give rise to different edge structures as shown in the bottom panels.}
\label{fig:schematic}
\end{figure*}

The hallmark signature of the SPT phases in SSH type lattices is the presence of the bulk-boundary correspondence that associates bulk topological invariants with certain zero-energy edge states in the system with open boundaries. For example the topological phase for an SSH chain exhibits two degenerate edge states localized at the two end sites of the lattice. These features of the open boundary system strictly depend on the choice of unit cells of the periodic system. While such a choice is limited in an SSH chain, coupled SSH chains platforms provide a versatile ground to achieve topological phases and the edge states by leveraging the dimensionality to construct different types of unit cells. For example, for the two-leg SSH ladder, a completely different scenario appears. It has been shown that when the ladder is in the topological phase, the edge states appear at the two end sites of any one of the legs when the OBC is achieved by vertically cutting the ladder. This highlights the enhanced flexibility to manipulate the edge states when one moves from purely 1D chain to ladder geometries. Apart from this flexibility, SSH ladder systems also provide freedom to define different unit cells as compared to one dimensional SSH chain  which can in principle be used to engineer different topological phases, phase transitions and edge states. Moreover, from the experimental point of view, efforts have already been made in recent years to observe the topological phase transitions in different SSH ladder geometries using ring resonators~\cite{hchen}, metamaterials~\cite{prodan} and in system of ultracold atoms in optical lattices~\cite{pan_prr}.

Motivated by these, in this work we explore possible route to construct different unit cells in a two-leg SSH ladder model to engineer different topological phases and their associated edge states. We start from the two possible configurations; (i) uniform dimerization and (ii) staggered dimerization as shown in the left and right panels, respectively, of Fig.~\ref{fig:schematic}. We show that if different types of unit cells are chosen (demarcated as blue and magenta boxes) for both the uniform and staggered dimerization cases, then an interesting topological landscapes emerge. We reveal that for the uniform dimerization case (left panel), for both the blue and magenta unit cell choices, the system exhibits a topological to topological phase transition through a gapless region as the rung hopping is varied. However, for the staggered dimerization case, a much richer scenario emerges. For the unit cell marked by the blue box, the system exhibits a topological to topological transition, and for the magenta unit cell, a trivial to topological phase transition as a function of the rung hopping. Unlike the uniform dimerization case, here the transitions occur through gap closing points instead of the gap closing region. We quantify these topological phases and transitions through the signatures from the bulk-boundary correspondence. Interestingly, we obtain that, while all the topological phases exhibit the standard bulk-boundary correspondence of a two-leg ladder that dictates two localized edge modes, in the case of staggered dimerization, the topological phase corresponding to the blue unit cell exhibits four edge states when the inter-leg coupling is weak. After obtaining these theoretical insights, we obtain the signatures of these topological phases using topoelectric circuit platforms. To this end, we first simulate the lattice Hamiltonian using an equivalent topoelectric circuit and then experimentally construct the circuit using appropriate inductors ($L$) and capacitors ($C$). By  measuring  the impedance and voltage responses in both simulation and experiment, we demonstrate the realization and engineering of edge states in SSH ladder systems.

\section{Model}\label{modelmethod}

We consider a system of non-interacting spinless fermions on a two-leg ladder, where each leg individually exhibits SSH model type hopping dimerization as depicted in Fig.~\ref{fig:schematic}. 
The model that describes such a system is given by the Hamiltonian
\begin{align}
H
= - \sum_{j} \Big[ t_a - (-1)^j \alpha_a \Big]
\left( a_j^\dagger a_{j+1} + \text{H.c.} \right)
\nonumber\\- \sum_{j} \Big[ t_b - (-1)^j \alpha_b \Big]
\left( b_j^\dagger b_{j+1} + \text{H.c.} \right)
+H_i
\label{eq:ham}.
\end{align}
where $H$ represents the Hamiltonian of non-interacting spinless fermions
\begin{align}
    H_i =  
    \begin{cases}
        - t_p \sum\limits_{j} \left( a_j^\dagger b_j + \text{H.c.} \right)~\text{for unit cell 1} \\
        - t_p \sum\limits_{j} \left( a_j^\dagger b_{j+1} + \text{H.c.} \right)~\text{for unit cell 2 and 3}
    \end{cases}
\end{align}
$a_j(a_j^\dagger)$ and $b_j(b_j^\dagger)$ are the particle annihilation (creation) operators on leg-a and leg-b, respectively, where $j$ represents the rung index. 
$t_a$ and $t_b$ are the hopping amplitudes along the legs and $\alpha$ represents the modulation parameter that dictates the hopping dimerization. 
$t_p$ represents the inter-leg hopping amplitude, which is also called the rung hopping. 

The ladder system possesses two distinct dimerization patterns. 
(i) Uniform dimerization configuration, where both the SSH legs of the ladder are in the same phase, being either topological or trivial, as illustrated in the left panel of Fig.~\ref{fig:schematic}, (ii) staggered dimerization configuration, where one leg of the ladder is in the topological phase while the other remains in the trivial phase, as shown in the right panel of Fig.~\ref{fig:schematic}.
To realize the two dimerization patterns discussed above, we choose the following parameter settings. 
The uniform dimerization configuration is obtained by taking equal hopping amplitudes along the two legs, $t_a = t_b = t$, and identical modulation strengths, $\alpha_a = \alpha_b = \alpha$. 
In contrast, the staggered dimerization configuration is implemented by choosing equal hopping amplitudes, $t_a = t_b = t$, but opposite modulation strengths, $\alpha_a = -\alpha_b = \alpha$. 
Throughout this work, we fix $t = 0.6$ and $\alpha = 0.4$, which yield the strong and weak bond amplitudes $t_1 = t + \alpha = 1.0$ and $t_2 = t - \alpha = 0.2$, respectively, along each legs of the ladder.

Due to the peculiarity of the ladder structure, it is possible to define different unit cells. For both dimerization patterns, three different choices of unit cells are marked as considered, as highlighted by the red, blue, and magenta boxes in Fig.~\ref{fig:schematic}(a) of both the left and right panels.  In the following we will show that each such choice of unit cell leads to different topological features.
We analyze the topological properties corresponding to all these configurations and demonstrate how such choices can lead to different edge state realizations under OBC. Then we obtain signatures of edge states from the electric circuit simulations. 
\begin{figure*}[t]
\centering
\includegraphics[width=1\linewidth]{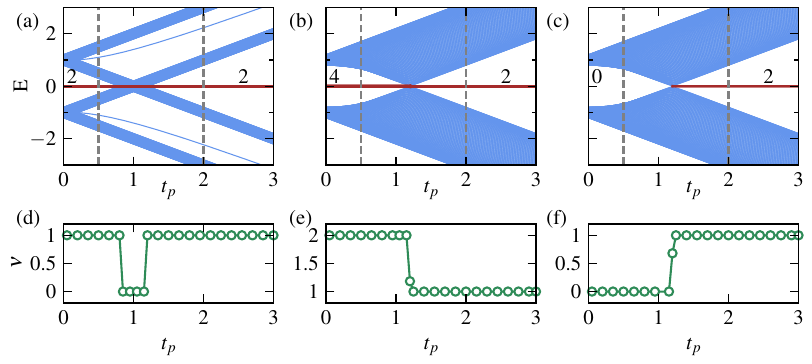}
\caption{Energy spectra of a two-leg ladder as a function of the inter-leg hopping strength $t_p$. Panel (a) shows the energy spectrum corresponding to the unit cell 2 (blue box) of Fig.~\ref{fig:schematic}(a).  Panels (b) and (c) show the energy spectrum for the unit cells 2 (blue box) and 3 (magenta box) of Fig.~\ref{fig:schematic}(b) corresponding to $(t_1,t_2)=(0.2,1.0)$ and $(1.0,0.2)$, respectively. The red central points indicate the zero energy edge states. The system size is $N=200$ sites. Panels (d)–(f) show the corresponding winding number $\nu/\pi$ as a function of $t_p$ for the parameter regimes shown in panels (a)–(c). The isolated states in (a) are the edge states in the high energy gaps. The numbers in the middle white regions (gaps) indicate the number of edge states.}
\label{fig:energy_spectrum}
\end{figure*}

\begin{figure*}[t]
\centering
\includegraphics[width=1\linewidth]{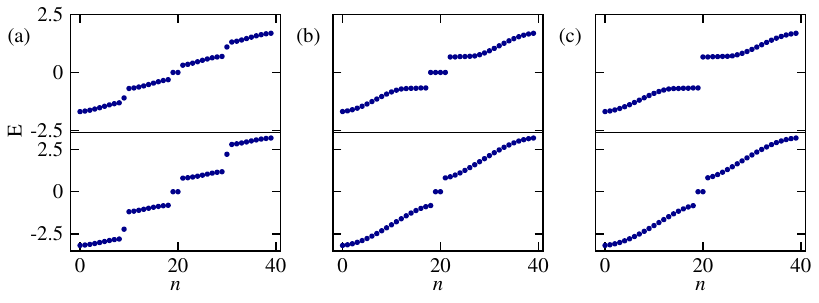}
\caption{The energy spectra $E$ as a function of the eigenstate index ($n$) for two representative cuts, $t_p=0.5$ and $t_p=2.0$ corresponding to gray dashed lines in Fig.~\ref{fig:energy_spectrum}(a)–(c). The isolated states appear at the zero energy within the gap (except for the upper panel of (c)) correspond to the edge states} 
\label{fig:edge_state_count}
\end{figure*}

\section{Results}
In this section we discuss our main results which shows unit-cell dependent topological phases and edge state engineering, their characterization from the topoelectric circuit simulation and then experimental observation. 

\subsection{Unit cell dependent topological phases}
We analyze the topological properties of the two-leg SSH ladder model by studying the real-space energy spectrum of the Hamiltonian shown in Eq.~\ref{eq:ham}  under OBC together with the corresponding bulk topological invariant for an equivalent periodic system. 
The topological properties of coupled SSH ladders under OBC, specifically when both chains are in the topological phase (a uniform dimerization case) or when one chain is topological and the other trivial (a staggered dimerization case), and are coupled via perpendicular interleg hopping, have been discussed previously by some of us in Ref.~\cite{padhan_ladder}. 
When both chains are topological, two pairs of zero-energy edge modes appear in the decoupled limit. 
However, upon turning on the perpendicular rung hopping, these modes hybridize and move away from zero energy, rendering the topology weak. 
For unit cell 1 with both chains initially in the trivial phase, no zero-energy edge states are present, and the bulk properties remain identical to the case where both chains are topological, as both configurations share the same bulk spectrum (not shown).
On the other hand, in the staggered dimerization case, unit cell 1 is known to exhibit a topological phase transition accompanied by the emergence of a pair of zero-energy edge states. 
A shift of this unit cell by one rung leaves the bulk Hamiltonian invariant but restores reflection symmetry about the horizontal axis. 
As a result, the bulk physics remains unchanged, while the location of edge states changes from one leg of the ladder to the other. This phenomenon has been discussed in detail in Ref. \cite{padhan_ladder}. 

In this work, we consider different unit cells  (2 and 3) as depicted in the Fig.~\ref{fig:schematic}. In the case of uniform dimerization (left panel of Fig.~\ref{fig:schematic}) the unit cells 2 and 3 are related by a reflection symmetry and are equivalent to each other. However, for the staggered dimerization case, such equivalence is absent. Therefore, we focus only on unit cell 2 for the former case and on 2 and 3 for the latter. 
The real-space energy spectra as functions of inter-leg hopping $t_p$ are shown in Fig.~\ref{fig:energy_spectrum}(a) for unit cell 2 of the uniform dimerization configuration and in Fig.~\ref{fig:energy_spectrum}(b) and (c) for unit cells 2 and 3, respectively, of the staggered dimerization configuration. 
All calculations are performed for a system of $N = 200$ sites, with fixed hopping amplitudes $t_1 = 1.0$ and $t_2 = 0.2$.
We obtain that for all the three cases, the system exhibits a gapped to gapped phase transition as a function of $t_p$ at half filling of the band (central white regions). While for the uniform dimerization case, the gapped to gapped transition occurs through a gap closing region (Fig.~\ref{fig:energy_spectrum}), for the staggered dimerization case, the transition occurs through a gap closing point ($t_p^c$) for both the unit cell configurations (Fig.~\ref{fig:energy_spectrum} b and c). This difference in the gap-closing scenarios can be understood from the band structure. For example the dispersion relation for the uniform dimerization case [Fig. \ref{fig:energy_spectrum}(a)] can be written as
\[
E(k)=\pm\left(|\rho(k)|\pm t_p\right),
\]
where
\[
|\rho(k)|=\sqrt{t_1^2+t_2^2+2t_1t_2\cos k}.
\]
The gap closing condition $t_p=|\rho(k)|$ suggests that for some momentum $k$, the bulk gap remains closed over the finite range
\begin{equation}
    |t_1-t_2|\le t_p \le |t_1+t_2|
\end{equation} 
for fixed values of $t_1$ and $t_2$. Hence, within this range, for every value of $t_p$, there exists a value of $k$ at which the gap vanishes, resulting in an extended gapless region. Similarly, for the staggered SSH ladder case [Fig.~\ref{fig:energy_spectrum}(b) and (c)],  the gap-closing condition is satisfied only at $t_p=t_1+t_2$ at $k=0$, and therefore the bulk gap vanishes only at a single critical point. Consequently, the topological transition in the staggered ladder occurs through an isolated gap-closing point separating two gapped phases, whereas the uniform ladder exhibits an extended gapless phase due to the identical hopping structure of the two legs. It is to be noted that, in all the cases the spectrum exhibits states at zero energies (marked as red) inside the gapped region(s).

\begin{figure}[t]
\begin{center}
\includegraphics[width=1.\columnwidth]{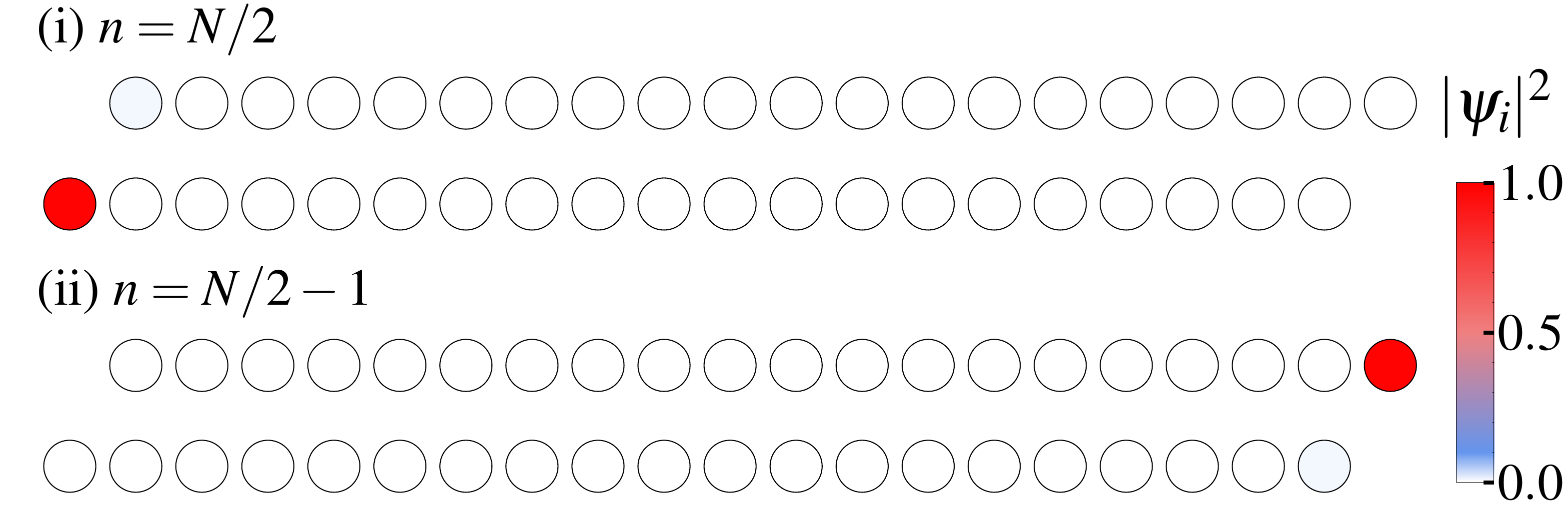}
\end{center}
\caption{On-site probability densities $|\psi_i|^2$ of two states located at the center of the spectrum for a system of size $N=40$ sites at $t_p = 2.0$. The maximum density at the edges for each case indicate the topological nature of the two states.}
\label{fig:two_edge_state}
\end{figure}

To identify the number of zero-energy states, we plot the real-space energy spectrum as a function of the site indices (i) for two representative cuts of $t_p$ in Fig.~\ref{fig:energy_spectrum}(a)-(c), i.e, $t_p=0.5$  and $2.0$  chosen within the gapped phases on either sides of the bulk gap closing 
which are shown in the upper and lower panels, respectively, of Fig.~\ref{fig:edge_state_count}(a)–(c). 
For the uniform dimerization case [Fig.~\ref{fig:edge_state_count}(a)], two zero-energy states persist on both sides of the gap closing. 
In contrast, for unit cell~2 of the staggered dimerization configuration [Fig.~\ref{fig:edge_state_count}(b)], four zero-energy states appear for  $t_p < t_p^c \sim 1.2$. However, in the gapped phase for $t_p > t_p^c \sim 1.2$, only two states with zero energy survive [Fig.~\ref{fig:edge_state_count}(b) (lower panel)]. 
For unit cell~3 [Fig.~\ref{fig:edge_state_count}(c)], no zero-energy states appear within the gap for  $t_p < t_p^c \sim 1.2$ (upper panel). However, beyond the gap closing point ($t_p > t_p^c \sim 1.2$), a pair of zero-energy states emerges in the gap (lower panel). The appearance of such zero-energy states within the bulk gap are typical signatures of edge states in topological phases.
\begin{figure}[t]
\begin{center}
\includegraphics[width=1.\columnwidth]{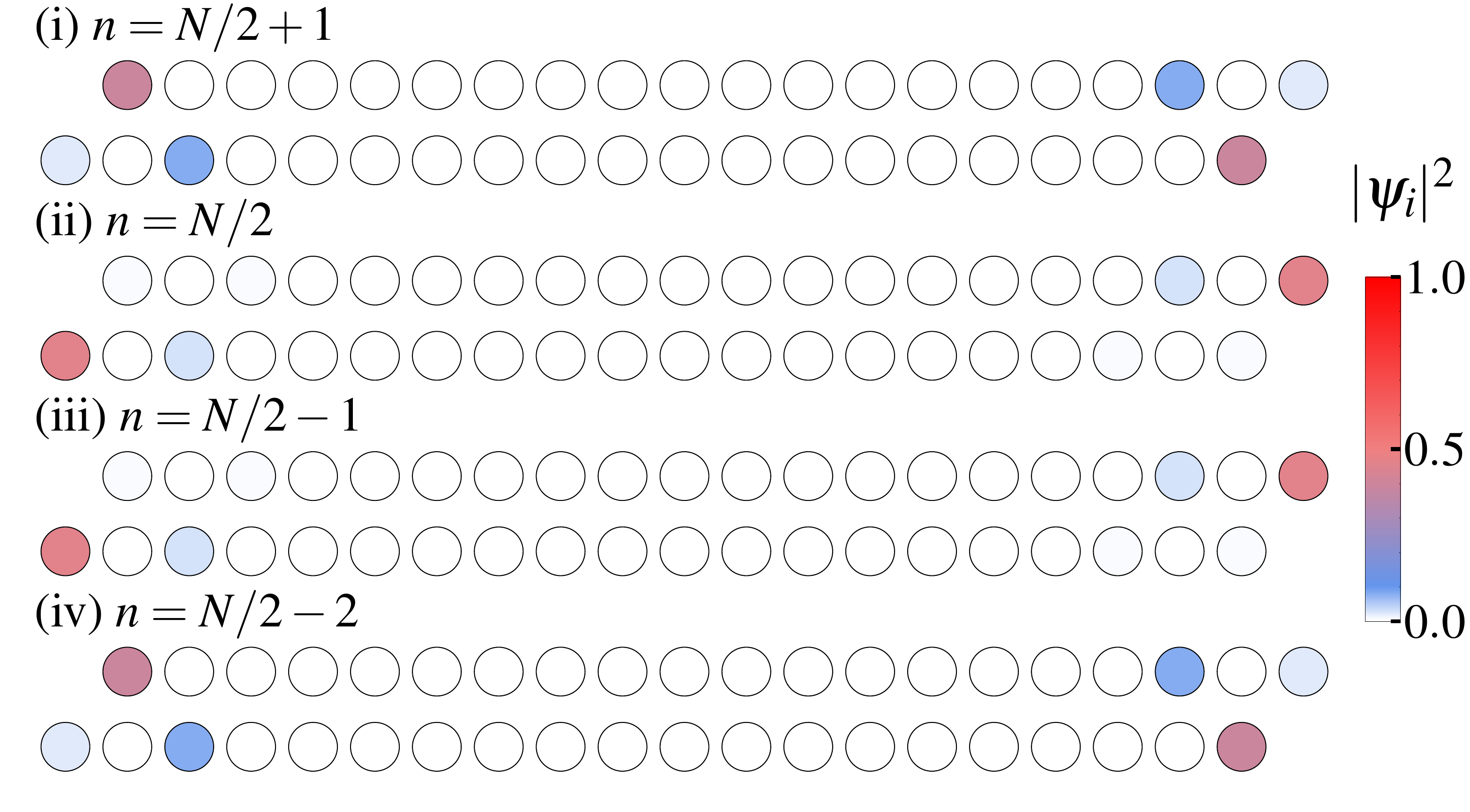}
\end{center}
\caption{On-site probability density $|\psi_i|^2$ of four states located at the center of the spectrum for a system of size $N=40$ sites at $t_p = 0.5$. The maximum density at the edges for each case indicate the topological nature of the four states.} \label{fig:four_edge_state}
\end{figure}
In order to characterize the zero-energy states appearing here, we calculate the on-site probability densities $|\psi_i|^2$ for a system of $N=40$ sites. For this demonstration, we only choose the case of unit cell 2 of the staggered dimerization case (Fig.~\ref{fig:energy_spectrum}(b)) and plot the onsite probability densities $|\psi_i|^2$ at each site corresponding to these zero-energy states for $t_p=0.5$ and $t_p=2.0$ in Fig.~\ref{fig:four_edge_state} and Fig.~\ref{fig:two_edge_state}, respectively. These also correspond to the states shown in the upper and the lower panels of Fig.~\ref{fig:edge_state_count}(b).
The lattice sites are represented by circles, and the face colors indicate the on-site probability densities. 
All the four zero energy states are clearly localized at the edge sites for $t_p=0.5$ as shown in Fig.~\ref{fig:four_edge_state}(i-iv), indicating that all of them are edge states of the system.
On the other hand, only two states are localized at the edge sites for $t_p=2.0$, which are shown in Fig.~\ref{fig:two_edge_state} (i) and (ii). This confirms the existence of only two edge states in the system. Similar scenarios are also found in other two cases of Fig.~\ref{fig:energy_spectrum} (not shown). 

\begin{figure}[b]
    \centering \includegraphics[width=1\linewidth]{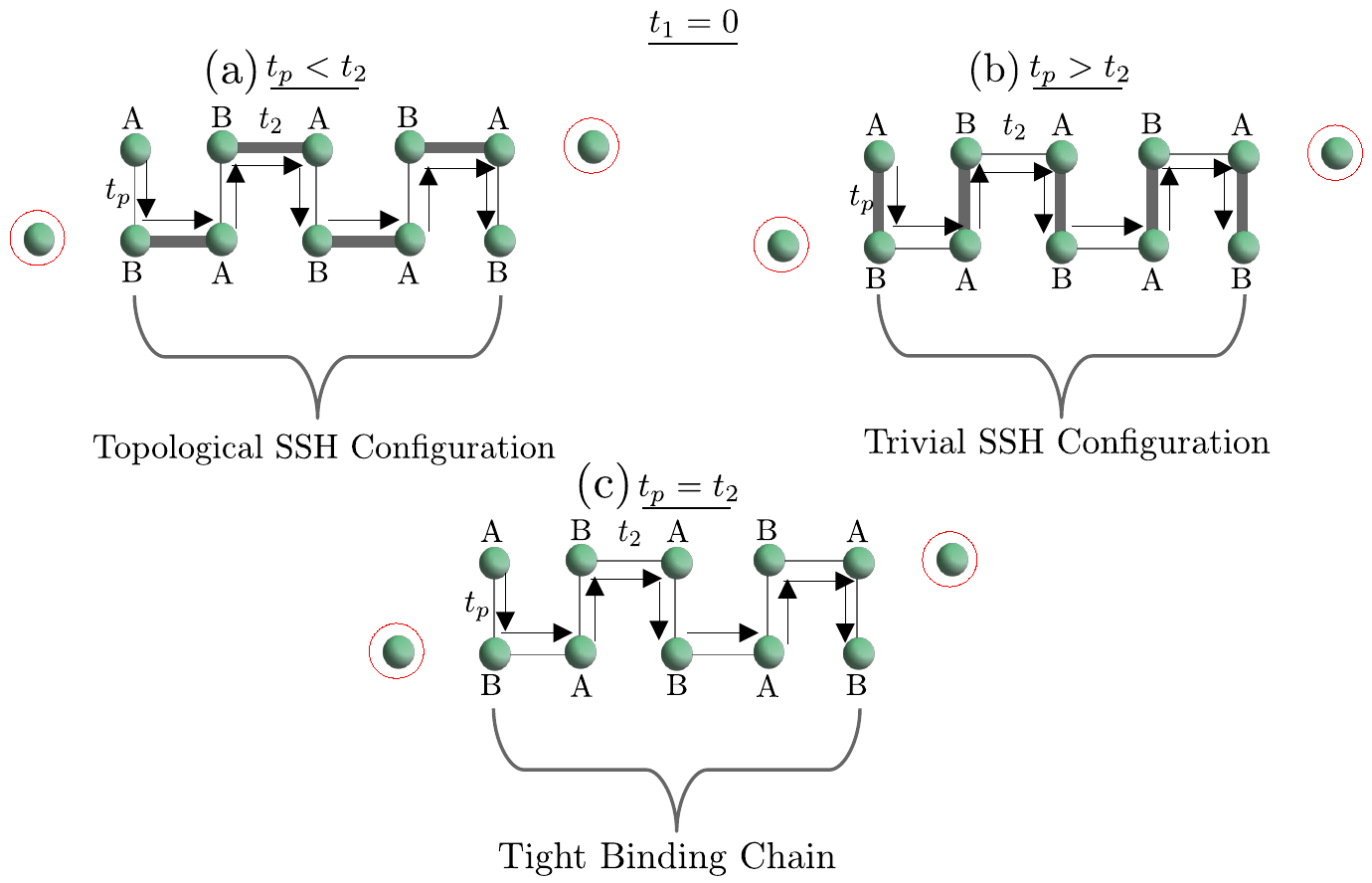}
    \caption{The schematic illustration of the physical origin of different numbers of edge states in different $t_p$ limits corresponding to Fig.~\ref{fig:energy_spectrum}(b). The red circles indicate the isolated boundary sites, while the arrows indicate the formed SSH chain. (a) For $t_p<t_2$, the SSH chain is topological and, together with the two isolated boundary sites, supports four zero-energy edge states. (b) For $t_p>t_2$, the SSH chain is trivial, leaving only the two isolated boundary states. (c) At $t_p=t_2$, the SSH chain reduces to a uniform tight-binding chain, while the isolated boundary sites remain decoupled.}
    \label{fig:cartoon}
\end{figure}

To gain further physical insight into the origin of the different edge-states in Fig.~\ref{fig:energy_spectrum}, we consider the limiting case $t_1=0$, illustrated schematically in Fig.~\ref{fig:cartoon}. In this limit, the physical origin of the edge states becomes transparent. As an illustrative example, we discuss the transition from four to two zero-energy edge states shown in Fig.~\ref{fig:energy_spectrum}(b). In the limit of $t_1=0$, the system decomposes into an SSH chain together with two isolated boundary sites highlighted using red circles. For $t_p<t_2$ [Fig.~\ref{fig:cartoon}(a)], the SSH chain formed from $t_p-t_2-t_p-t_2-...$ pattern is in its topological phase and supports two zero-energy edge states. Together with the two isolated boundary states, this situation results in a total of four zero-energy edge states.
At the critical point $t_p=t_2$ [Fig.~\ref{fig:cartoon}(c)], the SSH chain becomes a uniform tight-binding chain and undergoes the topological phase transition through a bulk gap closing. While the edge states originating from the topological SSH chain disappear at this critical point, the two isolated boundary sites remain completely decoupled from the bulk in the $t_1=0$ limit. Consequently, two zero-energy edge states persist even at the gapless transition point.
For $t_p>t_2$ [Fig.~\ref{fig:cartoon}(b)], the SSH chain enters the trivial phase and therefore no longer supports zero-energy edge states. As a result, the only remaining zero-energy edge states are those associated with the isolated boundary sites, yielding a total of two edge states. Although Fig.~\ref{fig:energy_spectrum} corresponds to finite $t_1$, this picture remains valid by adiabatic continuity as long as the bulk gap does not close. A finite $t_1$ merely hybridizes the isolated boundary sites weakly with the bulk and shifts the phase boundary quantitatively without changing the origin of the edge states.
A similar analysis can be carried out for the other edge-state transitions shown in Fig.~\ref{fig:energy_spectrum}, where the limiting configurations likewise reveal the origin of the corresponding edge modes.

Up to this point, it is understood that the system for different choices of the unit cells exhibits different gapped phases, and some of these gapped phases host zero-energy edge states. While these signatures hint towards the topological phases, to concretely characterize the topological nature of the bulk, we compute the Zak phase~\cite{berry_phase,Zak1989}, which serves as the relevant bulk topological invariant for the present one-dimensional system. 

The Zak phase is evaluated numerically as
\begin{equation}
  \nu = -\mathfrak{I}\ \ln \prod_{j=0}^{N-1} \text{det}(\langle u_{m,k_j} |u_{n,k_{j+1}} \rangle),
  \label{eq:berry_phase}
\end{equation}
where $\mathfrak{I}$ denotes the imaginary part. Here, $k_j = 2\pi j/(Na)$ represents the $j$th momentum point in the Brillouin zone, $a$ is the lattice spacing, and $|u_{m,k_j}\rangle$ denotes the cell-periodic part of the Bloch eigenstate corresponding to the occupied band $m$ at momentum $k_j$. 
The determinant is taken to properly account for the multiband nature of the occupied subspace. 
A trivial insulating phase is characterized by $\nu = 0$, whereas a topologically nontrivial phase corresponds to $\nu = \pi$.
The Zak phase, calculated as a function of $t_p$, is shown by green circles in the bottom panels of Fig.~\ref{fig:energy_spectrum}. 
For the uniform dimerization configuration, a gapped phase hosting a single pair of zero-energy states is characterized by $\nu/\pi = 1$. 
Upon increasing $t_p$, the bulk gap closes, rendering the Zak phase ill-defined. 
As the gap reopens at $t_p> t_p^c$, we get the Zak phase $\nu/\pi = 1$ which corresponds to the two zero energy edge states.
For unit cell 2 of the staggered dimerization configuration, we obtain  $\nu/\pi = 2$ which corresponds to two pairs of zero-energy edge states for  $t_p < t_p^c$. 
These states merge with the bulk at $t_p=t_p^c$ where the bulk gap closes. 
Upon gap reopening ($t_p > t_p^c$), a single pair of zero-energy edge states survives, accompanied by a reduction of the Zak phase to $\nu/\pi = 1$. 
In contrast, unit cell 3 of the staggered configuration exhibits a vanishing Zak phase at $t_p< t_p^c$ which matches with absence of any edge states. This confirms the trivial nature of the gapped phase in this region of $t_p$.  
However, increasing $t_p$ results in a finite Zak phase of $\nu/\pi = 1$ for $t_p > t_p^c$. This corresponds to the appearance of a pair of zero-energy edge states in the spectrum.
These results demonstrate a direct correspondence between the bulk Zak phase and the emergence of zero-energy mid-gap edge states, confirming the topological nature of the phases and phase transitions.

At this point, we want to comment that even though one chooses different unit cell configurations, the bulk property will remain the same, and hence the band structure.
All the bulk Hamiltonians are equivalent to each other up to a unitary transformation.
In the following, we explicitly show few of them.
Corresponding to the unit cell highlighted using the red box for both uniform and staggered configuration (in Fig.~\ref{fig:schematic}), the Fourier space Hamiltonian is given by
\begin{equation}
    {H_1}_{\text{u}} = \begin{pmatrix}
    0 & -t_p & 0 & \mu_1\\
    -t_p & 0 & \mu_1 & 0\\
    0 & \mu_1^* & 0 & -t_p\\
    \mu_1^* & 0 & -t_p & 0\\ 
    \end{pmatrix}\label{eq:uni_band_ham_1}
\end{equation}
and 
\begin{equation}
    {H_1}_{\text{s}} = \begin{pmatrix}
    0 & -t_p & 0 & \mu_2\\
    -t_p & 0 & \mu_3 & 0\\
    0 & \mu_3^* & 0 & -t_p\\
    \mu_2^* & 0 & -t_p & 0\\ \end{pmatrix}\label{eq:stag_band_ham_1}
\end{equation}
where we have set the lattice constant to unity, with $\mu_1=-t_2-t_1e^{-ik}$, $\mu_2=-t_1-t_2e^{-ik}$, and $\mu_3=-t_2-t_1e^{-ik}$ with $k$ representing the quasi-momentum.
Similarly, the Fourier space Hamiltonian corresponding to the unit cell highlighted by the blue box, corresponding to uniform and staggered configurations are given by
\begin{equation}
    {H_2}_{\text{u}} = \begin{pmatrix}
    0 & 0 & -t_p & \mu_3\\
    0 & 0 & \mu_2 & -t_pe^{-ik}\\
    -t_p & \mu_2^* & 0 & 0\\
    \mu_3^* & -t_pe^{ik} & 0 & 0\\ 
    \end{pmatrix}
    \label{eq:uni_band_ham_2}
\end{equation}
and
\begin{equation}
    {H_2}_{\text{s}} = \begin{pmatrix}
    0 & 0 & -t_p & \mu_1\\
    0 & 0 & \mu_1 & -t_pe^{-ik}\\
    -t_p & \mu_1^* & 0 & 0\\
    \mu_1^* & -t_pe^{ik} & 0 & 0\\ 
    \end{pmatrix}
    \label{eq:stag_band_ham_2}
\end{equation}
respectively.
Both ${H_1}_{\text{u}}$ and ${H_2}_{\text{u}}$ are equivalent to each other up to a unitary transformation given by ${U_{12}}_\text{u}{H_1}_{\text{u}}{U_{12}}_\text{u}^\dagger={H_2}_{\text{u}}$. Also, ${H_1}_{\text{s}}$ and ${H_2}_{\text{s}}$ are equivalent to each other up to a unitary transformation given by ${U_{12}}_\text{s}{H_1}_{\text{s}}{U_{12}}_\text{s}^\dagger={H_2}_{\text{s}}$, where
\begin{equation}
    {U_{12}}_\text{u}={U_{12}}_\text{s} = \begin{pmatrix}
    1 & 0 & 0 & 0\\
    0 & 0 & e^{-ik} & 0\\
    0 & 1 & 0 & 0\\
    0 & 0 & 0 & 1\\ 
    \end{pmatrix}
    \label{eq:unitary_operator}
\end{equation}
In an analogous manner, one can show that the bulk Hamiltonian associated with unit cell 3 (highlighted by the magenta box) is unitarily equivalent to the bulk Hamiltonians indicated by the red and blue boxes for the respective configurations. 
In the following, we discuss the LC circuit realization of the phenomena discussed above.
\begin{figure}[t]
\centering
\begin{tikzpicture}
\node[inner sep=0] (img)  {
\resizebox{\columnwidth}{!}{%
\begin{circuitikz}[scale=0.85, every path/.append style={line width=2pt}]
\ctikzset{line width=1.2pt}
\tikzset{
  every node/.append style={font=\Large},
  every label/.append style={font=\Large}
}

\def\dx{3}
\def\yu{4}
\def\yl{1}
\def\yg{-2}
\def\yt{7}

\draw (0,\yt) -- (17,\yt);
\draw (-\dx,\yg) -- (17,\yg);
\draw (17,\yg) -- (17,\yt);
\node[ground] at (17,\yg) {};

\tikzset{
  redbubble/.style={
    circle,
    fill=red,
    inner sep=3pt,
    minimum size=8pt
  }
}
\node[redbubble, label={[xshift=-3mm,yshift=0mm]below:{\Large A}}] at (0,\yu) {};
\node[redbubble, label={[xshift=0mm,yshift=0mm]below:{\Large B}}] at (5*\dx,\yu) {};
\node[redbubble, label={[xshift=0mm,yshift=0mm]above:{\Large C}}] at (-\dx,\yl) {};
\node[redbubble, label={[xshift=3mm,yshift=0mm]above:{\Large D}}] at (4*\dx,\yl) {};

\draw[draw=red]
(0,\yu) to[C,l=\textcolor{red}{$C_1$}] (\dx,\yu);

\draw[draw=blue]
(\dx,\yu) to[C,l=\textcolor{blue}{$C_2$}] (2*\dx,\yu);

\draw[draw=red]
(2*\dx,\yu) to[C,l=\textcolor{red}{$C_1$}] (3*\dx,\yu);

\draw[draw=blue]
(3*\dx,\yu) to[C,l=\textcolor{blue}{$C_2$}] (4*\dx,\yu);

\draw[draw=red]
(4*\dx,\yu) to[C,l=\textcolor{red}{$C_1$}] (5*\dx,\yu);

\draw[draw=red]
(-\dx,\yl) to[C,l=\textcolor{red}{$C_1$}] (0,\yl);

\draw[draw=blue]
(0,\yl) to[C,l=\textcolor{blue}{$C_2$}] (\dx,\yl);

\draw[draw=red]
(\dx,\yl) to[C,l=\textcolor{red}{$C_1$}] (2*\dx,\yl);

\draw[draw=blue]
(2*\dx,\yl) to[C,l=\textcolor{blue}{$C_2$}] (3*\dx,\yl);

\draw[draw=red]
(3*\dx,\yl) to[C,l=\textcolor{red}{$C_1$}] (4*\dx,\yl);

\foreach \x in {0,\dx,2*\dx,3*\dx,4*\dx} {
  \draw[draw=violet]
  (\x,\yu) to[C,l=\textcolor{violet}{$C_3$}] (\x,\yl);
}

\foreach \x in {\dx,2*\dx,3*\dx,4*\dx} {
  \draw (\x,\yu) to[L,l=$L$, line width=0.6pt] (\x,\yt);
}
\foreach \x in {0,\dx,2*\dx,3*\dx} {
  \draw (\x,\yl) to[L,l=$L$, line width=0.6pt] (\x,\yg);
}

\draw (0,\yu) -- (0,\yu+0.5);

\draw
(0,\yu+0.5) -- (-0.5,\yu+0.5)
          to[C,l={$C_2$}] (-0.5,\yu+2.5)
          -- (0,\yu+2.5);

\draw
(0,\yu+0.5) -- (0.5,\yu+0.5)
          to[L,l_=$L$, line width=0.6pt] (0.5,\yu+2.5)
          -- (0,\yu+2.5);

\draw (0,\yu+2.5) -- (0,\yt);

\draw (5*\dx,\yu) -- (5*\dx,\yu+0.5);

\draw
(5*\dx,\yu+0.5) -- (5*\dx-0.5,\yu+0.5)
           to[L,l=$L$, line width=0.6pt] (5*\dx-0.5,\yu+2.5)
           -- (5*\dx,\yu+2.5);

\draw
(5*\dx,\yu+0.5) -- (5*\dx+0.5,\yu+0.5)
           to[C,l_={\rotatebox[origin=c]{-90}{$C_2+C_3$}}] (5*\dx+0.5,\yu+2.5)
           -- (5*\dx,\yu+2.5);

\draw (5*\dx,\yu+2.5) -- (5*\dx,\yt);

\draw (-\dx,\yl) -- (-\dx,\yl-0.5);

\draw
(-\dx,\yl-0.5) -- (-\dx-0.5,\yl-0.5)
          to[C,l_={$C_2+C_3$}] (-\dx-0.5,\yl-2.5)
          -- (-\dx,\yl-2.5);

\draw
(-\dx,\yl-0.5) -- (-\dx+0.5,\yl-0.5)
          to[L,l=$L$, line width=0.6pt] (-\dx+0.5,\yl-2.5)
          -- (-\dx,\yl-2.5);

\draw (-\dx,\yl-2.5) -- (-\dx,\yg);

\draw (4*\dx,\yl) -- (4*\dx,\yl-0.5);

\draw
(4*\dx,\yl-0.5) -- (4*\dx-0.5,\yl-0.5)
           to[L,l_=$L$, line width=0.6pt] (4*\dx-0.5,\yl-2.5)
           -- (4*\dx,\yl-2.5);

\draw
(4*\dx,\yl-0.5) -- (4*\dx+0.5,\yl-0.5)
           to[C,l={$C_2$}] (4*\dx+0.5,\yl-2.5)
           -- (4*\dx,\yl-2.5);

\draw (4*\dx,\yl-2.5) -- (4*\dx,\yg);

\draw
(-3,\yu) to[sV,l=$I_{\mathrm{AC}}$] (-3,\yt)
         -- (0,\yt);
\draw (-3,\yu) -- (0,\yu);

\end{circuitikz}
}
};
\end{tikzpicture}
\caption{Schematics of the SSH ladder topolectrical circuit in a staggered dimerization configuration. Each unit cell consists of three capacitors, $C_1$, $C_2$, and $C_3$, connected by identical inductors $L$ between neighboring capacitors. The circuit is driven by an AC current source with amplitude $I_{\text{AC}}$. Unit-cell configurations 2 and 3 are realized by tuning the capacitance ratio $C_1/C_2$ to values smaller than or larger than unity. Nodes specified by A, B, C, and D are the target nodes where we measure the impedance.}
\label{fig:circuit_schematic}
\end{figure}

\subsection{Topoelectric realization of the edge states}
After getting insights on the topological phases and various edge states properties, in this section we observe the signatures of the topological phases  by obtaining the signatures of the zero energy edge states using topoelectric circuits. Topo-electric circuits have recently emerged as a very simple, yet powerful and flexible platform for simulating topological phenomena in tight-binding lattices~\cite{band_structure_topo_ckt}. The peculiar similarity between the quantum lattice model and the circuit Laplacian provides an interesting avenue to map the Hamiltonian matrix to the admittance matrix. As a result the eigenstates of the quantum model can be viewed as the voltage response across different nodes. 
These systems offer several key advantages, including precise control over parameters, direct access to edge responses via impedance measurements, and the ability to simulate both Hermitian and non-Hermitian topological phases~\cite{majorana_non_hermitian, NHSE_topo_ckt, reciprocal_se_topo_ckt, gbz_nh_topo_ckt, unconventional_skin_mode, topo_ckt_nhse,maity_aubry_andre, topo_defect_engeneering, topo_corner_mode_ckt, non_hermitian_topo_sensor, loc_ctrl_of_defect_states,dipendu_rony}. Due to such properties, topo-electric circuits have recently attracted a great deal of attention to observe topological phases in various geometries~\cite{topoelectric_theory_bitan}. 
Importantly, circuit realizations of SSH-type models have demonstrated clear signatures of topological boundary modes, providing an experimentally accessible route to visualize bulk-boundary correspondence in real space~\cite{ching_hua_jalil}.

\begin{figure*}[t]
\begin{center}
\includegraphics[width=2.\columnwidth]{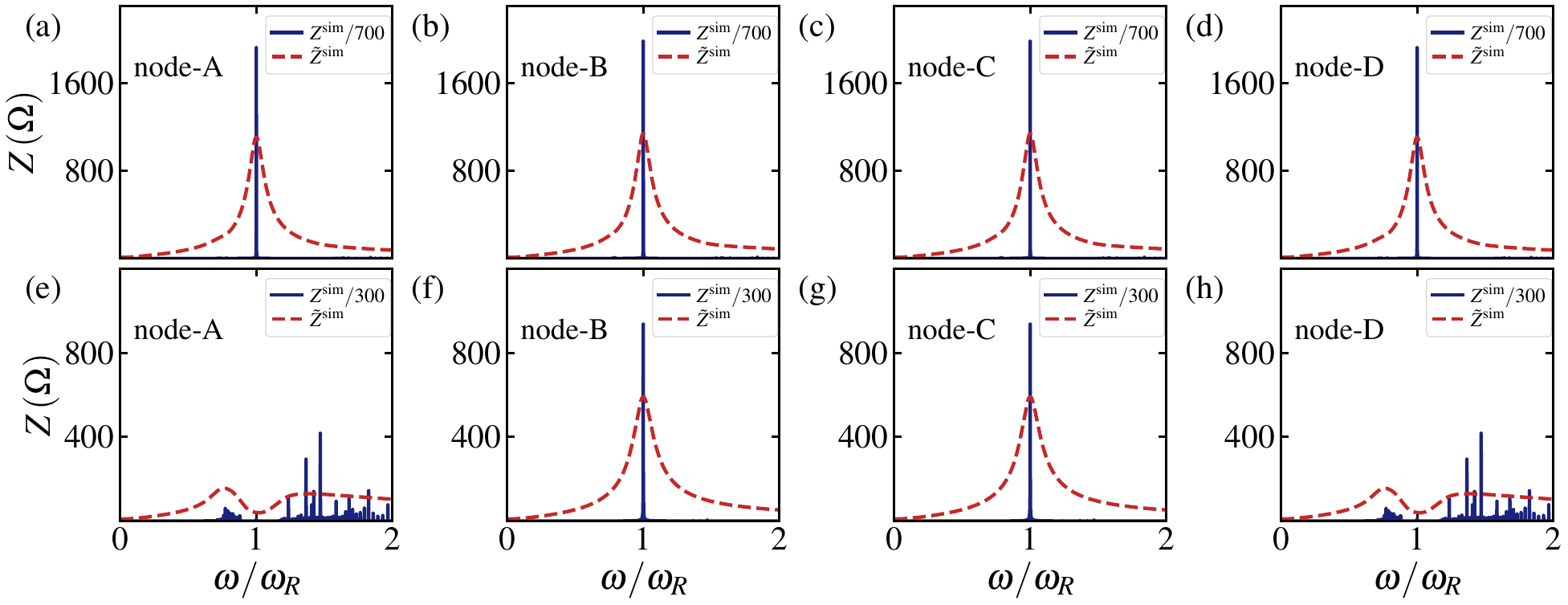}
\end{center}
\caption{The simulated two-point impedances $Z^\text{sim}$ (blue solid curves) measured across all the target edge nodes  as a function of normalized AC frequency $\omega/\omega_R$ for a circuit made out of $50$ unit cells ($N=200$ sites). The circuit parameters are set to $C_1=100 nF,~C_2=470 nF$, and $L= 10mH$. (a-d) Show the impedances for $C_3=100 nF$ and (e-h) for $C_3=1000nF$. The simulated two point impedance with internal resistance ($\tilde{Z}^{\text{sim}}$) shown as red dashed curves computed by using internal resistances of $10 \Omega$ added to $10mH$ inductors, $30 \Omega$ to $100nF$ capacitors, $2.5 \Omega$ to $470nF$ capacitors and $1 \Omega$ to $1000 nF$ capacitors in series.  The blue curves are appropriately scaled for better visibility. }\label{fig:impedance_peak}
\end{figure*}
In what follows, we will first simulate and then perform experiments to obtain the signatures of topological edge states using LC circuits. 
We consider the situation shown in Fig.~\ref{fig:energy_spectrum}(b) to capture the topological features under OBC, which corresponds to the unit cell configuration 2 in Fig.~\ref{fig:schematic} (right panel). The lattice model for this configuration under OBC (Fig.~\ref{fig:schematic}(b2)) can be mapped into a circuit model as depicted in Fig.~\ref{fig:circuit_schematic}. Although for the actual simulation purposes we consider larger systems, we use a smaller circuit of $5$ unit cells for the following discussion. The hopping dimerization in each leg $t_1$ and $t_2$ can be achieved by considering capacitors with alternating capacitances $C_1$ and $C_2$, respectively. The inter-leg coupling $t_p$ is achieved by using another capacitor of capacitance $C_3$.  The junction between the capacitors in each legs are connected by inductors $L$ which are grounded. In addition, we also add extra capacitors to the edge nodes to match their onsite potentials with the bulk nodes as depicted in Fig.~\ref{fig:circuit_schematic}. Following Kirchoffs's law, the eigen equation corresponding to the above circuit can be written as 
\begin{equation}
\mathbf{I}=(\mathcal{L}+W)\mathbf{V}\equiv J\mathbf{V},
\end{equation}
where $\mathbf{V}$ and $\mathbf{I}$ are vectors composed of the node voltages and injected nodal currents. 

Here the grounded Laplacian $J$ is the sum of the circuit Laplacian $\mathcal{L}$, which encodes the network structure, and the diagonal matrix ($W$) specifies the grounding configuration, which in our case is given by
\begin{equation}
\begin{aligned}
W &= i\omega\,\mathrm{Diag}\Big(
C_2-\frac{1}{\omega^2 L}, -\frac{1}{\omega^2 L}, -\frac{1}{\omega^2 L},
-\frac{1}{\omega^2 L}, -\frac{1}{\omega^2 L}, \\
&\qquad C_1+C_2-\frac{1}{\omega^2 L},
C_1+C_2-\frac{1}{\omega^2 L},
-\frac{1}{\omega^2 L}, -\frac{1}{\omega^2 L}, \\
&\qquad -\frac{1}{\omega^2 L},
-\frac{1}{\omega^2 L},
C_2-\frac{1}{\omega^2 L}
\Big).
\end{aligned}
\end{equation}
The circuit Laplacian 
\begin{equation}
\mathcal{L}=D-C,
\end{equation}
with $C$ and $D$ being the adjacency and the degree matrices, respectively (See Appendix \ref{app:appendix_a}).
With these ingredients in hand, we derive the associated Laplacian to our grounded circuit as\\

\resizebox{\columnwidth}{!}{%
\(
J =i\omega
\left(
\begin{array}{cccccccccccc}
D & -C_1 & 0 & 0 & 0 & ... & -C_3 & 0 & 0 & 0 & 0 \\
-C_1 & D & -C_2 & 0 & 0 & ... & 0 & -C_3 & 0 & 0 & 0 \\
0 & -C_2 & D & -C_1 & 0 & ... & 0 & 0 & -C_3 & 0 & 0 \\
0 & 0 & -C_1 & D & -C_2 & ... & 0 & 0 & 0 & -C_3 & 0 \\
0 & 0 & 0 & -C_2 & D & ... & 0 & 0 & 0 & 0 & -C_3 \\
0 & 0 & 0 & 0 & -C_1 & ... & 0 & 0 & 0 & 0 & 0 \\
0 & 0 & 0 & 0 & 0 & ... & -C_1 & 0 & 0 & 0 & 0 \\
-C_3 & 0 & 0 & 0 & 0 & ... & D & -C_2 & 0 & 0 & 0 \\
0 & -C_3 & 0 & 0 & 0 & ... & -C_2 & D & -C_1 & 0 & 0 \\
0 & 0 & -C_3 & 0 & 0 & ... & 0 & -C_1 & D & -C_2 & 0 \\
0 & 0 & 0 & -C_3 & 0 & ... & 0 & 0 & -C_2 & D & -C_1 \\
0 & 0 & 0 & 0 & -C_3 & ... & 0 & 0 & 0 & -C_1 & D 
\end{array}
\right)
\)%
}\\

\noindent
with $D=C_1+C_2+C_3-\frac{1}{\omega^2 L}$.
Now setting $D=0$, we get the resonant frequency associated with the circuit as 
\begin{equation}
    \omega_R=\frac{1}{\sqrt{L(C_1+C_2+C_3)}}.
    \label{eq:res_freq}
\end{equation}
Using the above circuit configuration, we probe the topological signatures through impedance measurements using LTspice simulations~\cite{topoelectric_theory_bitan,topo_ckt_thomale,active_topo_ckt,bic_topo_ckt,Izmailian_2014,Izmailian_two_point}.
We measure the two-point impedance~\cite{two_point_resistance,theory_of_impedance} given by $Z_{mn}=(V_m-V_n)/I$, where $m$ and $n$ represent the nodes among which the voltage difference is being calculated, and $I$ represents the externally injected AC current flowing from node-$m$ to node-$n$.
A pronounced peak in the impedance indicates that the admittance matrix possesses at least one eigenvalue that is close to zero. 
In the circuit realization of the topologically non-trivial phases, such a strong response is a signature of gapless boundary modes associated with a topologically nontrivial phase~\cite{impedance_responses}. 
To identify the emergence of edge modes for the configuration shown in Fig.~\ref{fig:circuit_schematic}, we inject current at the four target nodes marked as A, B, C, and D at the edges in Fig.~\ref{fig:circuit_schematic} and measure the voltage difference with respect to the ground.

For the case under consideration, we expect to observe a phase transition from a topological phase possessing four edge states to another topological phase with two edge states.
To obtain the signatures of four edge states, which occurs in the limit of weak inter-leg coupling, we choose the circuit parameters as 
$C_1 = 100\,n\mathrm{F}$, 
$C_2 = 470\,n\mathrm{F}$, 
$C_3 = 100\,\,n\mathrm{F}$, 
and $L = 10\,\mathrm{mH}$ and plot the two point impedance $Z$ as a function of normalized angular frequency $\omega/\omega_R$ in the upper panel of Fig.~\ref{fig:impedance_peak}. With the choice of the above parameters, the computed resonance frequency using Eq.~\ref{eq:res_freq} is obtained as $f = \omega_R/2\pi=1.944\,\mathrm{kHz}$. Our simulation shows prominent impedance peaks at nodes A, B, C, and D, indicating the presence of four zero-energy modes. 
For the case of strong inter-leg coupling, we choose $C_1 = 100\,n\mathrm{F}$, 
$C_2 = 470\,n\mathrm{F}$, 
$C_3 = 1\,\mu\mathrm{F}$, 
and $L = 10\,\mathrm{mH}$ and plot $Z$ in units of $K\Omega$ as a function of normalized angular frequency $\omega/\omega_R$ in the lower panel of Fig.~\ref{fig:impedance_peak}.
Now, the resonance frequency shifts to $f = 1.270\,\mathrm{kHz}$. 
In this regime, two impedance peaks are observed only at nodes B and C, whereas no peaks are observed at nodes A and D.
\begin{figure}[b]
\begin{center}
\includegraphics[width=1.\columnwidth]{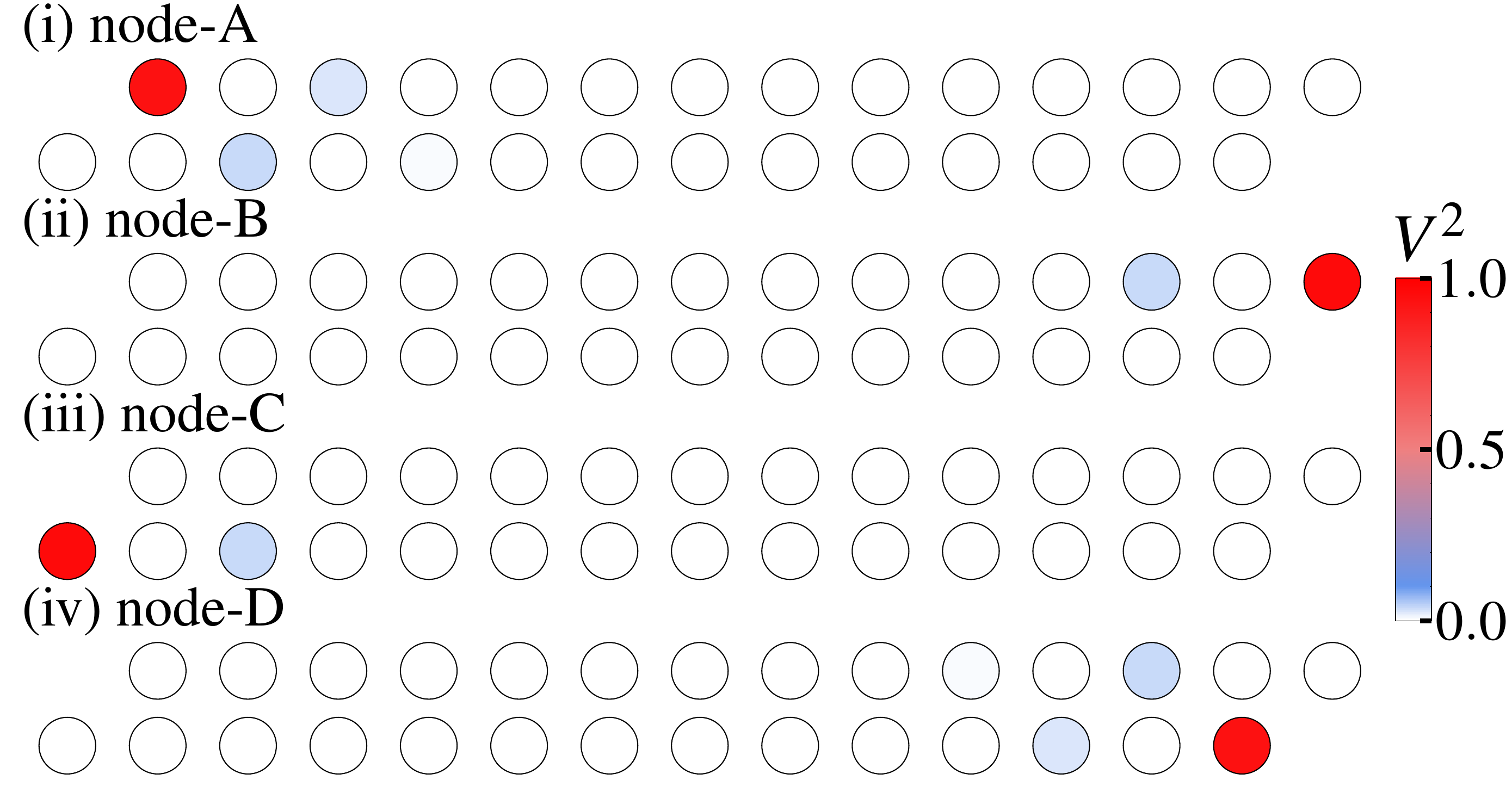}
\end{center}
\caption{Depicts the simulated data for local voltage distribution $V^2$ of all four edge excitations with the circuit parameters set to $C_1=100n F,~C_2=470n F$,~$C_3=100n F$ and $L= 10mH$ .}
\label{fig:four_edge_state_lc_ckt}
\end{figure}
\begin{figure}[t]
\begin{center}
\includegraphics[width=1.\columnwidth]{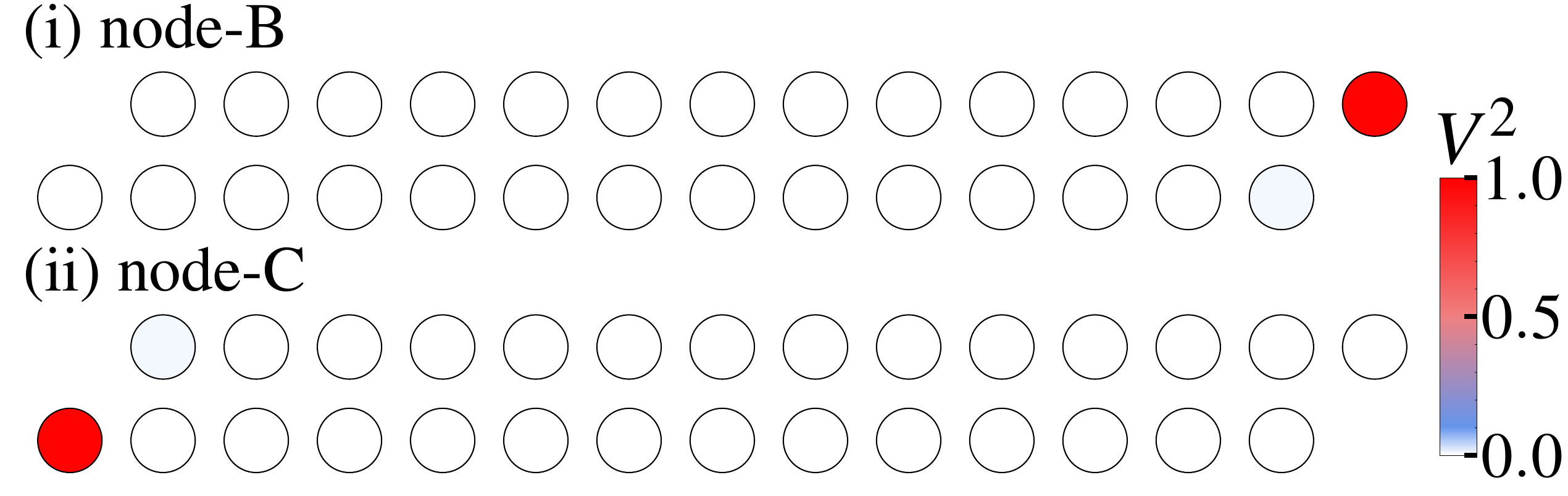}
\end{center}
\caption{Depicts the simulated data for local voltage distribution $V^2$ of all two edge excitations with the circuit parameters set to $C_1=100n F,~C_2=470n F$,$C_3=1000n F$ and $L= 10mH$ .} 
\label{fig:two_edge_state_lc_ckt}
\end{figure}
To further complement these observations, we probe the voltage response ($V$) which is equivalent to the probability amplitudes of the tight binding model given in Eq.~\eqref{eq:ham}.
\begin{figure}[b]
\begin{center}
\includegraphics[width=1.\columnwidth]{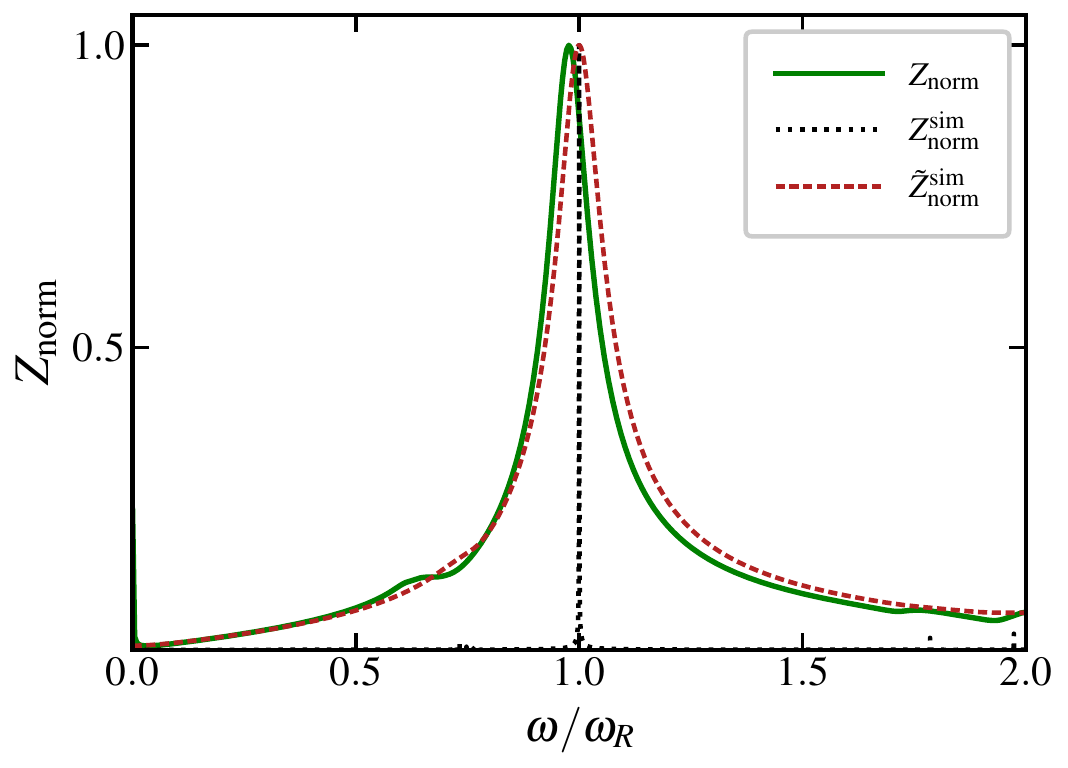}
\end{center}
\caption{Measured two-point normalized impedance $Z_{\text{norm}}$ (green solid curve) as a function of $\omega/\omega_R$ for an SSH chain constructed on a breadboard using circuit elements $C_1=100nF$, $C_2=470nF$, $L=10mH$ and total sites $N=14$. Simulation data for an equivalent circuit without (with ) internal resistance $Z_{\text{norm}}^{\text{sim}}$ ($\tilde{Z}_{\text{norm}}^{\text{sim}}$) is shown as a black dotted curve (red dashed curve). $\tilde{Z}_{\text{norm}}^{\text{sim}}$ is calculated by using resistances of $30\Omega,~2.5\Omega~\text{and}~10\Omega$ in series with $C_1$, $C_2$ and $L$, respectively.}
\label{fig:1D_SSH_exp}
\end{figure}

To this end, we apply a sinusoidal excitation of frequency $f = 1.944\,\mathrm{kHz}$, and measure the voltage response across all the nodes which are plotted in Fig.~\ref{fig:four_edge_state_lc_ckt}. 
The voltage distribution revealed strong localization of the zero energy mode at all the terminal nodes, which agrees well with the numerical prediction that there exist four edge states in this parameter regime.
On the other hand, when excited at $f = 1.270\,\mathrm{kHz}$, the voltage distribution shows localization at the two edge nodes as shown in Fig.~\ref{fig:two_edge_state_lc_ckt}.
The simulation results show excellent agreement with theoretical predictions.

\subsection{Experimental observation of the topological edge states}
\begin{figure*}[!t]
\begin{center}
\includegraphics[width=1.5\columnwidth]{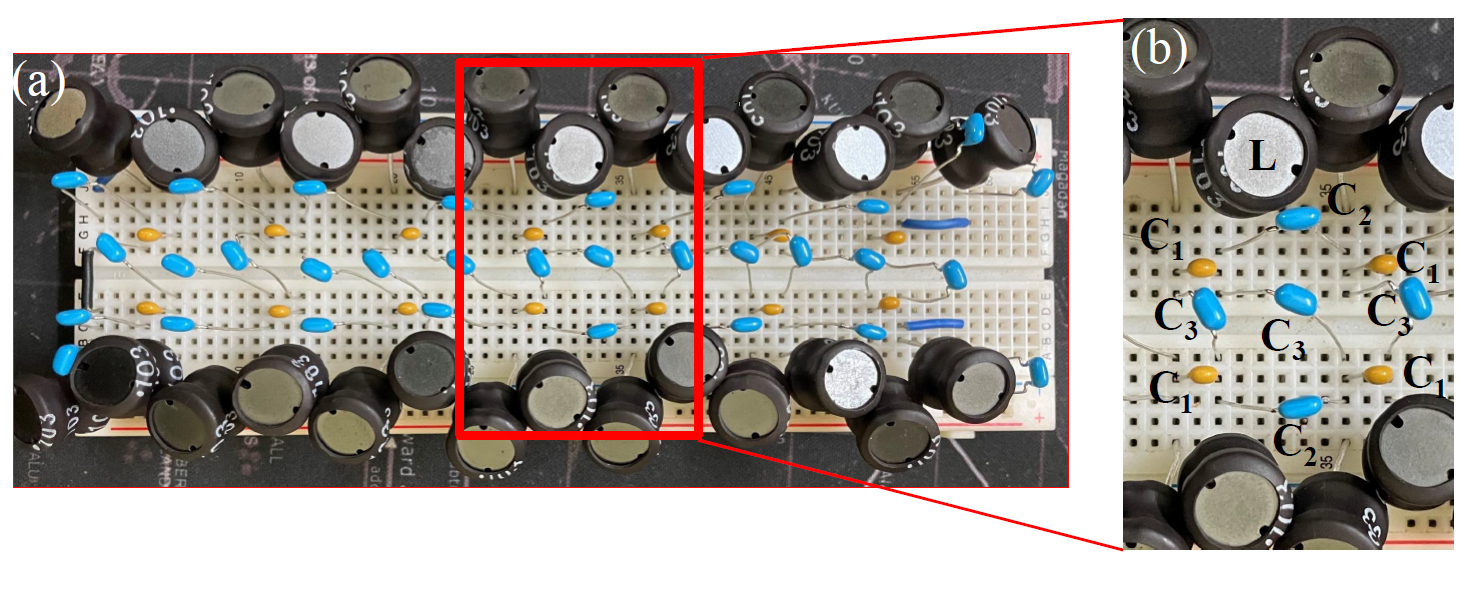}
\end{center}
\caption{(a) The breadboard construction of the SSH ladder with seven unit cells corresponding to the schematic shown in Fig.~\ref{fig:circuit_schematic}. (b) The zoomed in portion shows the position of the individual components. } 
\label{fig:bb_image}
\end{figure*}

We then perform real experiments by constructing the circuits to observe the signatures of the edge states. To this end, we design the circuit shown in Fig.~\ref{fig:bb_image} on a breadboard to mimic a system of size $N=28$ sites. While the breadboard circuits are known to be noisy due to the presence of unavoidable contact resistances along with internal resistances of inductors and capacitors, we demonstrate that for our purpose it captures essential physics. 

\begin{figure*}[!t]
\begin{center}
\includegraphics[width=2.\columnwidth]{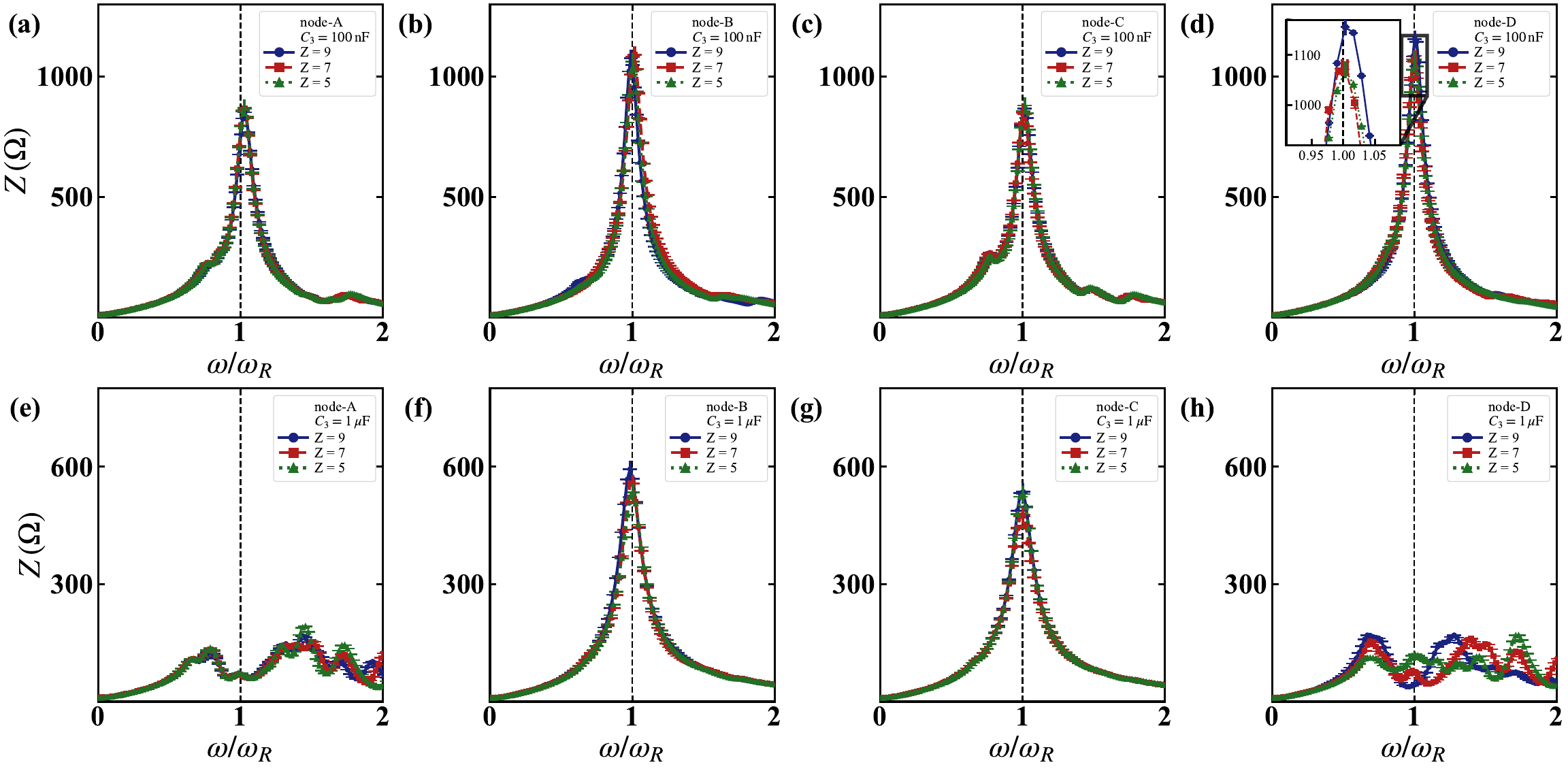}
\end{center}
\caption{Impedance $Z(\Omega)$ measured across all the target edge nodes (A, B, C and D) as a function of normalized AC frequency $\omega/\omega_R$ for circuits made out of $5,~7$ and $9$ unit cells ($N=20,~28$ and $36$ nodes respectively). (a-d) Impedance data for $C_1=100nF,~C_2=470nF,~C_3=100nF$ and $L= 10mH$ and (e-h) for $C_1=100nF,~C_2=470nF,~C_3=1000 n F$ and $L= 10mH$. The peaks at the resonance frequency in (a-d) and (e-h) indicate the presence of the edge states. The error bars are obtained by taking five observations for each system sizes. For some cases the error bars are smaller than symbol size. Inset of (d) shows the zoomed in portion of the main panel for better resolution as an exemplary scenario.} \label{fig:impedance_peak_exp}
\end{figure*}
Before moving on to the two-leg  SSH ladder model, we first perform the experiment by designing the 1D SSH chain using LC circuits on a breadboard to verify the accuracy of the platform. We construct an SSH chain of $N=14$ nodes, using inductors with $L=10 mH$  and capacitors $C_1=100 nF$ and $C_2=470 nF$ that satisfies the topological configuration of the model. We measure the two-point impedance $Z$ across the edge node as a function of frequency $f$ in the range between $1~Hz$ to $5~kHz$ using a chemical impedance analyzer (Model- IM 3590, HIOKI), with the AC excitation voltage maintained at $1V$ and plot the normalized impedance $Z_{\text{norm}}$ (green solid line)  as a function of $\omega/\omega_R$ in Fig.~\ref{fig:1D_SSH_exp}. Note that for the elements considered here, the resonant frequency $f=2.108 \,\mathrm{kHz}$ for the SSH circuit. A pronounced peak at $\omega/\omega_R=0.991$ corresponding to the circuit clearly suggests the presence of a zero-energy edge mode in the system. For comparison, we also plot the LTSpice simulation data for $Z_{\text{norm}}^{\text{sim}}$ (black dotted line) in Fig.~\ref{fig:1D_SSH_exp}.

While the peak position is consistent with the simulation data, the experimental impedance curve is more wider than that of the simulation data. This discrepancy can be attributed to the contact resistance of the breadboard circuit and the internal resistance of the circuit elements. To demonstrate this we repeat the simulation by adding internal resistances of each element used for the experiment to the circuit and plot the impedance data $\tilde{Z}_{\text{norm}}^{\text{sim}}$ (red dashed line) in Fig.~\ref{fig:1D_SSH_exp} as a function of $\omega/\omega_R$. While we could not include the effect of the contact resistance of the breadboard, a clear broadening of the impedance curve confirms the effect of internal resistance in the circuit elements. 
Another interesting thing to note is that although, the components used in our experiment are not ideal ones, we still get essential topological features. In particular, the capacitors in hand are of varying capacitances within a very small range around the desired values. However, such small variations do not affect the topological features in the circuit.

\begin{figure}[t]
\begin{center}
\includegraphics[width=1.\columnwidth]{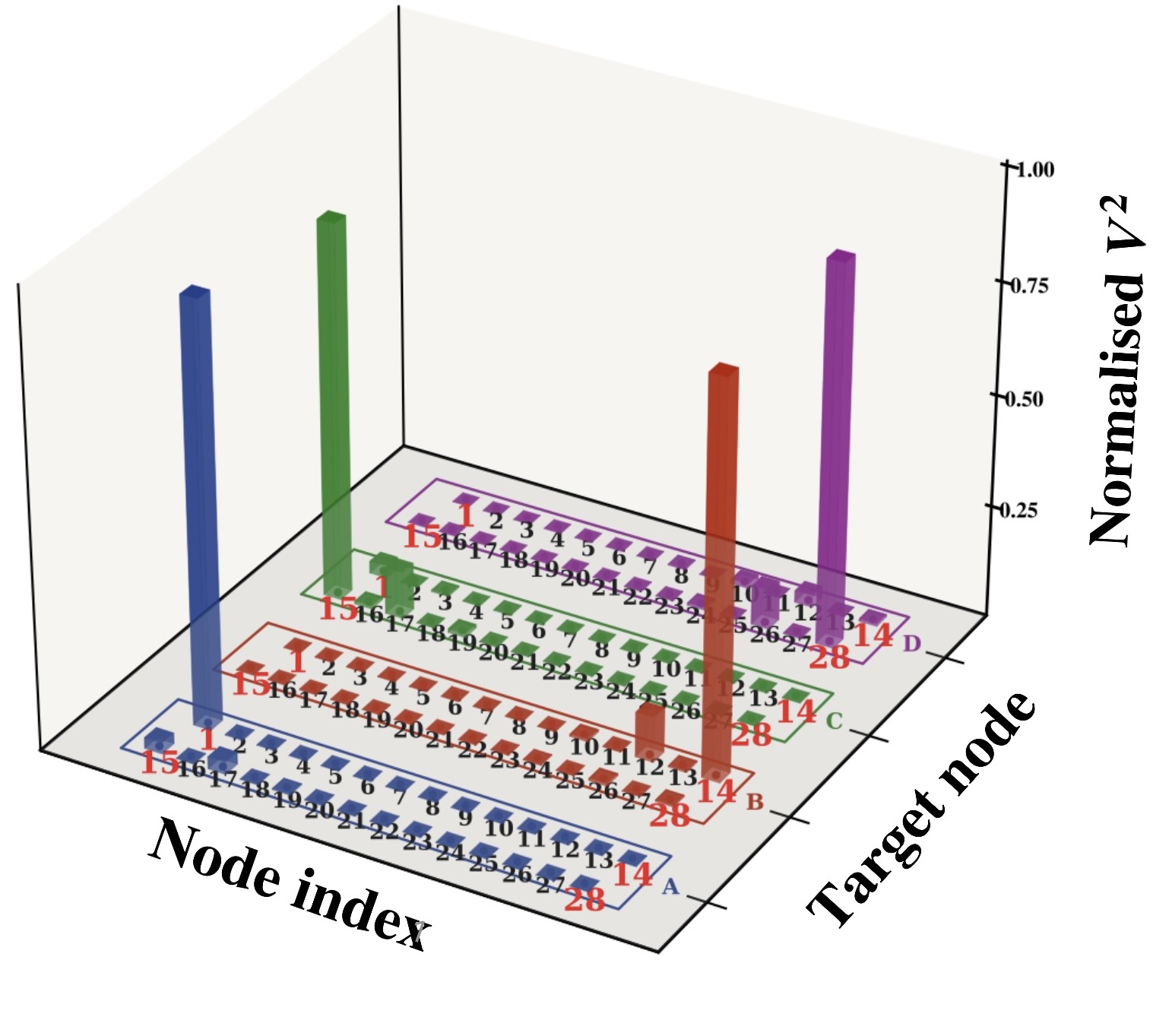}
\end{center}
\caption{Depicts the $V^2$ distribution across all nodes for all four edge excitations with the circuit component values $C_1=100nF,~C_2=470nF,~C_3=100nF$ and $L= 10mH$ which correspond to four topological edge states in the limit of weak inter-leg coupling (compare upper panel of Fig.~\ref{fig:impedance_peak_exp}).}
\label{fig:four_edge_state_exp}
\end{figure}

With this validation of the set-up in hand, we construct the SSH ladder for the circuit shown in Fig.~\ref{fig:circuit_schematic} using appropriate capacitors and inductors as depicted in Fig.~\ref{fig:bb_image} (see figure for details). 
We measure the two-point impedance $Z(\Omega)$ at the target edge nodes of the circuit shown in Fig.~\ref{fig:bb_image} (which correspond to the nodes located at A, B, C and D of Fig.~\ref{fig:circuit_schematic}) as functions of $\omega/\omega_R$ in which are shown in Fig.~\ref{fig:impedance_peak_exp}. The upper and lower panels  demonstrate the appearance of four and two edge states for weak and strong inter-leg couplings,  respectively. The clear peaks in Fig.~\ref{fig:impedance_peak_exp}(a-d) and Fig.~\ref{fig:impedance_peak_exp}(f-g) at $\omega/\omega_R\sim1$ indicate the appearance of the edge states, whereas absence of any peaks in  Fig.~\ref{fig:impedance_peak_exp}(e) and (h) indicate the absence of edge states. Note that in this case also we observe a slight deviation of the peak position from the expected value of $\omega/\omega_R=1$ and the broader peaks are due to parasitic resistance in the circuit as already discussed in the context of Fig.~\ref{fig:1D_SSH_exp}. To demonstrate the effect of parasitic resistance behind the broadened peaks, we have also plotted the simulated normalized impedance $\tilde{Z}^{\text{sim}}$ in Fig.\ref{fig:impedance_peak} for each target node by using appropriate resistances (see Fig.\ref{fig:impedance_peak} for details).  

\begin{figure}[h]
\begin{center}
\includegraphics[width=1.\columnwidth]{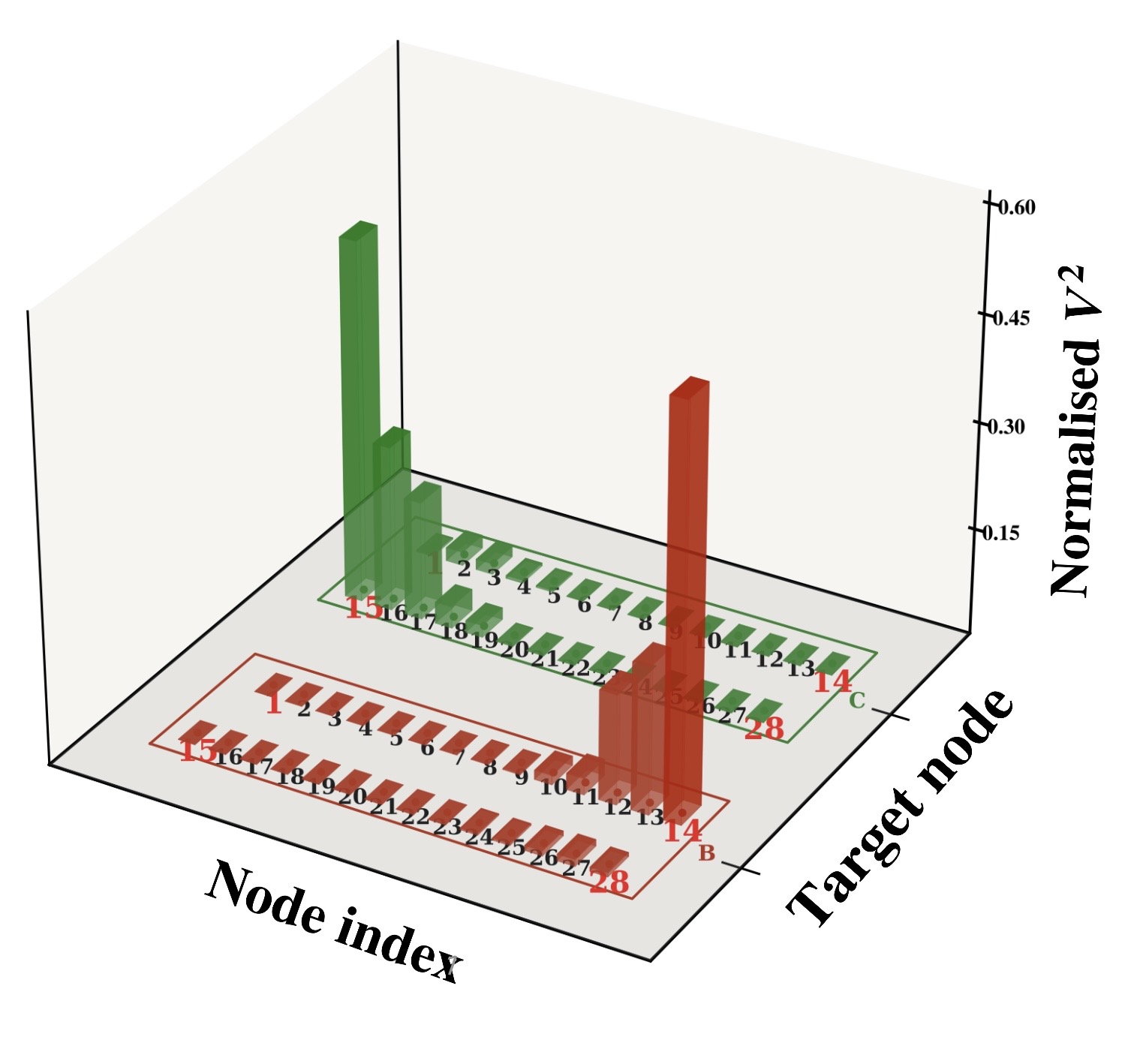}
\end{center}
\caption{Depicts the $V^2$ distribution across all nodes for all two edge excitations with the circuit component values $C_1=100nF,~C_2=470nF,~C_3=1000n F$ and $L= 10mH$ which correspond to two topological edge states in the limit of strong inter-leg coupling (compare lower panel of Fig.~\ref{fig:impedance_peak_exp}) .}
\label{fig:two_edge_state_exp}
\end{figure}

We also perform voltage measurements $V$ at each node of the circuit for the parameters considered in Fig.~\ref{fig:four_edge_state_lc_ckt} and Fig.~\ref{fig:two_edge_state_lc_ckt} and plot $V^2$ as a function of node index in Fig.~\ref{fig:four_edge_state_exp} and Fig.~\ref{fig:two_edge_state_exp} for the weak and strong inter-leg coupling regimes, respectively. Distinct peak in $V^2$ at the edge node for each case clearly complement the impedance peak shown in Fig.~\ref{fig:impedance_peak_exp}. 

The above analysis confirms the transition from a topological phase hosting four edge states to another topological phase hosting two edge states as the inter-leg coupling is tuned for the unit cell configuration 2. Similar analysis can also be performed for the other unit cell configurations to obtain the topological signatures (not shown). 

\section{Conclusion}\label{conclusion}
In this work, we investigated a two-leg SSH ladder and demonstrated the emergence of tunable topological phases, phase transitions, and edge states by appropriately defining the unit-cell configuration. We showed that different unit-cell configurations result in various topological phases and transition between them by tuning the inter-leg coupling. These topological phases exhibit essential bulk-boundary correspondence, allowing for possible edge state engineering. 
We proposed a topoelectric circuit simulations to observe these signatures through impedance measurements and the voltage response of the circuit. 
The existence of topological edge states was verified through impedance signatures and voltage distribution profiles obtained from circuit analysis via LTspice simulations. In the end we performed experiments using simple breadboard circuit arrangements and successfully obtained the signatures of the topological edge states. 

These findings reveal the flexibility in manipulating the unit-cells in a two-leg ladder model to obtain various topological phases and corresponding edge states. Moreover, the study highlights the potential of topoelectrical circuits as experimentally accessible platforms to realize lattice topological phases and also opens up avenues for investigating topological phases in disordered systems, higher-dimensional circuit architectures, and non-Hermitian environments. 

\section{Acknowledgment}
We thank Biswajit Paul and Soumya Ranjan Padhi for useful discussions. We thank Satyaprasad P Senanayak and Kanha Ram Khator for extending the laboratory facility for some of the measurements. T.M. acknowledges support from Science and Engineering Research Board (SERB), Govt. of India, through project No. MTR/2022/000382 and STR/2022/000023.

\appendix
\section{Additional Details}
\label{app:appendix_a}

We design the electrical circuit from the weighted graph where we take each vertices as circuit nodes and weighted edges as linear circuit components with edge weights as admittance of that circuit components. While we choose capacitors as edges in our circuit, one is free to choose inductive edges which does not change the overall circuit features. The weighted adjacency matrix $C$ is the matrix of admittances 
corresponding to our circuit, and the matrix elements can be defined as 
\[
C_{ij} = 
\begin{cases}
Y_{ij} & \text{if vertices i and j are connected, } i\ne j\\
0 & \text{otherwise,}
\end{cases}
\]  
where $Y_{ij}$ denotes the admittance of the circuit element connecting nodes $i$ and $j$. 
For the three types of couplings in our circuit, the admittances are given by
$i\omega C_1,~i\omega C_2,~ \text{and} ~i\omega C_3$,
depending on the corresponding capacitor in the network. 
So, the adjacency matrix for our circuit with $3$ types of couplings made by analogy with the tight-binding lattice will be,\\

\resizebox{\columnwidth}{!}{%
\(
C =i\omega
\left(
\begin{array}{cccccccccccc}
0 & C_1 & 0 & 0 & 0 & 0 & 0 & C_3 & 0 & 0 & 0 & 0 \\
C_1 & 0 & C_2 & 0 & 0 & 0 & 0 & 0 & C_3 & 0 & 0 & 0 \\
0 & C_2 & 0 & C_1 & 0 & 0 & 0 & 0 & 0 & C_3 & 0 & 0 \\
0 & 0 & C_1 & 0 & C_2 & 0 & 0 & 0 & 0 & 0 & C_3 & 0 \\
0 & 0 & 0 & C_2 & 0 & C_1 & 0 & 0 & 0 & 0 & 0 & C_3 \\
0 & 0 & 0 & 0 & C_1 & 0 & 0 & 0 & 0 & 0 & 0 & 0 \\
0 & 0 & 0 & 0 & 0 & 0 & 0 & C_1 & 0 & 0 & 0 & 0 \\
C_3 & 0 & 0 & 0 & 0 & 0 & C_1 & 0 & C_2 & 0 & 0 & 0 \\
0 & C_3 & 0 & 0 & 0 & 0 & 0 & C_2 & 0 & C_1 & 0 & 0 \\
0 & 0 & C_3 & 0 & 0 & 0 & 0 & 0 & C_1 & 0 & C_2 & 0 \\
0 & 0 & 0 & C_3 & 0 & 0 & 0 & 0 & 0 & C_2 & 0 & C_1 \\
0 & 0 & 0 & 0 & C_3 & 0 & 0 & 0 & 0 & 0 & C_1 & 0 
\end{array}
\right)
\)%
}\\

\noindent
\\
From the coupling matrix $C$, we now define the degree matrix 
$D$, which contains the total admittances from each node towards the rest of the circuit and is a diagonal matrix whose diagonal entries are given by the total weight connected to each vertex,
$$D_{ii} = \sum_{i=1}^N C_{ij} $$
and $D_{ij}=0 \text{, for } i \ne j$.
The degree matrix for this weighted graph will be a diagonal matrix and is given by\\

\resizebox{\columnwidth}{!}{%
\(
D = i\omega
\left(
\begin{array}{cccccccccccc}
C' & 0 & 0 & 0 & 0 & 0 & 0 & 0 & 0 & 0 & 0 & 0 \\
0 & C & 0 & 0 & 0 & 0 & 0 & 0 & 0 & 0 & 0 & 0 \\
0 & 0 & C & 0 & 0 & 0 & 0 & 0 & 0 & 0 & 0 & 0 \\
0 & 0 & 0 & C & 0 & 0 & 0 & 0 & 0 & 0 & 0 & 0 \\
0 & 0 & 0 & 0 & C & 0 & 0 & 0 & 0 & 0 & 0 & 0 \\
0 & 0 & 0 & 0 & 0 & C_1 & 0 & 0 & 0 & 0 & 0 & 0 \\
0 & 0 & 0 & 0 & 0 & 0 & C_1 & 0 & 0 & 0 & 0 & 0 \\
0 & 0 & 0 & 0 & 0 & 0 & 0 & C & 0 & 0 & 0 & 0 \\
0 & 0 & 0 & 0 & 0 & 0 & 0 & 0 & C & 0 & 0 & 0 \\
0 & 0 & 0 & 0 & 0 & 0 & 0 & 0 & 0 & C & 0 & 0 \\
0 & 0 & 0 & 0 & 0 & 0 & 0 & 0 & 0 & 0 & C & 0 \\
0 & 0 & 0 & 0 & 0 & 0 & 0 & 0 & 0 & 0 & 0 & C' 
\end{array}
\right)
\)%
}\\

\noindent where $C=C_1+C_2+C_3$ and $C'=C_1+C_3$.
As bulk vertices are connected to two intra-chain neighbors and one inter-chain neighbor, their diagonal entries are in the form of $D_{ii}=i\omega (C_1+C_2+C_3)$. In contrast, boundary vertices have fewer connections, leading to fewer contribution terms because of the open boundary condition chosen.

The matrices $C$ and $D$ define the graph Laplacian, $\mathcal{L}=D-C$, which forms the target operator to be implemented in the electrical circuit and is given as .\\

\resizebox{\columnwidth}{!}{%
 \(
 \mathcal{L} =i\omega
 \left(
 \begin{array}{cccccccccccc}
 C' & -C_1 & 0 & 0 & 0 & 0 & 0 & -C_3 & 0 & 0 & 0 & 0 \\
 -C_1 & C & -C_2 & 0 & 0 & 0 & 0 & 0 & -C_3 & 0 & 0 & 0 \\
 0 & -C_2 & C & -C_1 & 0 & 0 & 0 & 0 & 0 & -C_3 & 0 & 0 \\
 0 & 0 & -C_1 & C & -C_2 & 0 & 0 & 0 & 0 & 0 & -C_3 & 0 \\
 0 & 0 & 0 & -C_2 & C & -C_1 & 0 & 0 & 0 & 0 & 0 & -C_3 \\
 0 & 0 & 0 & 0 & -C_1 & C_1 & 0 & 0 & 0 & 0 & 0 & 0 \\
 0 & 0 & 0 & 0 & 0 & 0 & C_1 & -C_1 & 0 & 0 & 0 & 0 \\
 -C_3 & 0 & 0 & 0 & 0 & 0 & -C_1 & C & -C_2 & 0 & 0 & 0 \\
 0 & -C_3 & 0 & 0 & 0 & 0 & 0 & -C_2 & C & -C_1 & 0 & 0 \\
 0 & 0 & -C_3 & 0 & 0 & 0 & 0 & 0 & -C_1 & C & -C_2 & 0 \\
 0 & 0 & 0 & -C_3 & 0 & 0 & 0 & 0 & 0 & -C_2 & C & -C_1 \\ 
 0 & 0 & 0 & 0 & -C_3 & 0 & 0 & 0 & 0 & 0 & -C_1 & C' 
 \end{array}
 \right)
 \)%
 }\\
 
 \noindent
 The Laplacian $\mathcal{L}$ is independent of any choice of grounding and corresponds to a floating network.  \\
 
To define absolute node voltages and impedance measurement, the circuit must be grounded. All the grounding terms will appear at the diagonal of $J$. Grounding through inductors will give us a choice to cancel the on-site admittance terms at a particular frequency, just like how we set on-site potentials to be zero, we will call that frequency the resonant frequency. For this cancellation to happen, we need extra capacitors to ground from edge nodes, as they lack some connections due to finite chain boundaries. The resulting circuit is as already shown in  Fig.~\ref{fig:circuit_schematic}.
\paragraph*{}
\bibliography{references.bib}
\end{document}